\documentclass[aps,prl,twocolumn,showpacs,amsmath,amssymb]{revtex4}
\usepackage{graphicx}
\usepackage{dcolumn}
\usepackage{bm}
\begin{document}

\title{A ring accelerator for matter-wave solitons.}
\author{Alicia V. Carpentier and Humberto Michinel}
\affiliation{\'Area de \'Optica, Facultade de Ciencias de Ourense,\\ 
Universidade de Vigo, As Lagoas s/n, Ourense, ES-32004 Spain.}

\begin{abstract}
We describe a new technique to accelerate and control a confined cloud
of ultracold neutral atoms. In our configuration, a Bose-Einstein Condensate
trapped in a ring is accelerated by using an amplitude-modulated optical potential. 
The spreading of the atomic cloud is avoided by an adequate tuning of the 
scattering length. We also discuss other variations that offer new possibilities 
of robust control of atomic solitons. Our numerical results show that this cold 
matter accelerator is easily accessible in the frame of current experiments.
\end{abstract}


\maketitle
{\em Introduction.-} The control of neutral atomic beams is a challenging problem in physics
due to its potential applications in multiple fields like atom interferometry\cite{interferometer}, 
coherent control\cite{control}, atom clocks\cite{clock} or quantum information\cite{QI}, 
between others. The most common tool used to exert forces in neutral atoms are radiative 
interactions\cite{phillips} 
that allow to exchange momentum between photons and atoms, as in the case of radiative
pressure. Another good example is the dipole force, which appears in a laser intensity 
gradient and which can be interpreted in terms of absorption-stimulated emission cycles\cite{dipole}.
These processes allow to trap and cool atoms and are the basis of fundamental experiments
with ultracold atoms like Bose-Einstein Condensation (BEC) in alkalii gases\cite{anderson95}.

In the present work we will describe a simple technique that can be used to accelerate
an ultracold beam of neutral atoms stored in a ring reservoir\cite{reservoir}. 
The first of these storage devices was the {\em nevatron} \cite{Nevatron}, which consists of 
a simple double wire magnetic structure. After the nevatron, other schemes have been developed. 
In \cite{gupta} it was proposed a circular magnetic waveguide created by four 
coaxial circular electromagnets. In \cite{arnold} the condensate storage ring 
was obtained by using the quadrupole field created by a four-wire geometry. 
In all these models, the condensate moves in regular cycles with constant angular 
velocity and the cloud spreads along the ring during beam propagation. 
The method we propose overcomes the previous problems. It allows to increase the angular 
velocity of a BEC confined in a ring and to avoid the spreading of the cloud. Thus, the 
final result is a compact beam of ultracold neutral atoms that can be accelerated indefinitely
with the only limitation of the lifetime of the BEC. 

The trick is two-fold: in first place, we propose the use of a {\em static} amplitude-modulated 
optical potential.  The idea is similar to a previous proposal\cite{hecker} showing how to 
accelerate a condensate by employing a {\em moving} lattice, that transmits momentum 
to the condensate in packets of $2k\hbar$ (being $k$ the wavevector of the optical lattice). 
In second place, to avoid the spreading of the condensate during propagation and to get 
a suitable control of its motion, the scattering length has to be tuned to adequate negative 
values. The final result is a compact atomic soliton with increasing velocity in each round. The maximum final speed
that can be achieved depends on the lifetime of the condensate which limits the number 
of cycles. We have also simulated the effect of frequency-modulation 
(FM) lattices in order to show another phenomena like controllable localization of 
the cloud. In all the cases, our numerical calculations show that the parameters we 
use fit with usual values of current experiments.

Thus, we will first introduce the model and equations that we have used in this work, then
we will present the numerical simulations showing acceleration of a beam in one and two-dimensional 
configurations with AM lattices. We will also include some analytical calculations that corroborate
the numerical results. Finally, we will show the effect of FM lattices that allow to control 
the position of the condensate.

\begin{figure}[htb]
{\centering \resizebox*{1\columnwidth}{!}{\includegraphics{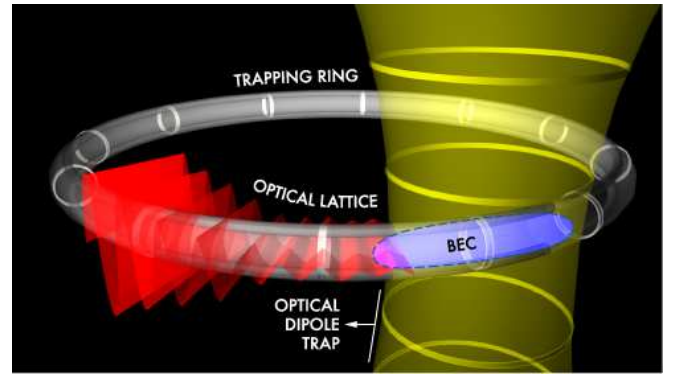}} \par}
\caption{[Color Online] Sketch of the system showing the trapping ring 
and the optical lattice (in red) used to accelerate the atomic soliton. 
The dipole trap (in yellow) is used to retain the cloud before it is 
accelerated.}
\label{fig1}
\end{figure}


{\em System and modeling equations.-} A sketch of our system is plotted in 
Fig. \ref{fig1}. It consists of a trapping ring illuminated with a periodic 
optical lattice. The mathematical description of the dynamics of the cloud 
is given by a Gross-Pitaevskii (GP) equation of the form
\begin{equation}
\label{GPE}
i \hbar \frac{\partial \Psi}{\partial t} =
- \frac{\hbar^2}{2 m} \nabla^{2}\Psi +
V(\vec{r})\Psi + U(\vec{r})|\Psi|^2 \Psi,
\end{equation}
where $\Psi$ is the order parameter, normalized to the number $N$ 
of atoms in the cloud: $N = \int |\Psi|^2 \ d^3 \mathbf{r},$
$U(\vec{r}) = 4 \pi \hbar^2 a(\vec{r})/m$ characterizes the 2-body interaction 
determined by the value of the scattering length $a$. In our model, we consider a 
ring potential $V(\vec{r})$ of radius $r_0$ and thickness $2L_\perp$ which varies 
azimuthally in the following form:
\begin{subequations}
 \begin{equation}
\label{toy}
V(r,\theta) = \begin{cases} V_r-V_0f(\theta) & r_0-L_\perp\leq  r \leq r_0+L_\perp \\
0 &  \text{elsewhere}
\end{cases}
\end{equation}
being
\begin{equation}
f(\theta) = \begin{cases} 
1 & -\alpha\leq \theta\leq 0\\
A\left(\theta\right)\sin^2\left(\nu \theta\right) &0\leq \theta \leq \beta\\
0 & \beta\leq\theta\leq 2\pi-\alpha
\end{cases}
\end{equation}
\end{subequations}

\begin{figure}[htb]
{\centering \resizebox*{1\columnwidth}{!}{\includegraphics{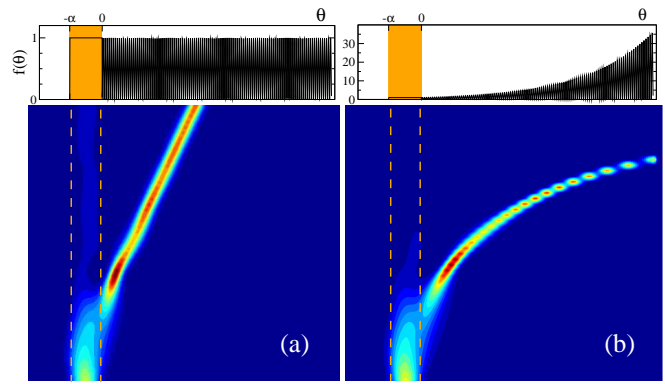}} \par}
\caption{[Color Online] One-dimensional simulation showing the acceleration 
of an atomic soliton. The cloud is emitted from the trap at $t=0s$
when the scattering length is switched to zero in the shaded region of the upper plots
(between dotted lines in lower images). In a) the amplitude of the lattice is constant, 
yielding to uniform motion. In b) the lattice is modulated with an exponential growth, 
obtaining the acceleration of the BEC. The horizontal axis corresponds to a path 
$l=200L_\perp$. The width of the cloud is $w\approx L=20L_\perp$. In the bottom pictures, 
the vertical axis is time from $t=0$ to $t=1.5s$.}
\label{fig2}
\end{figure}

Thus, the trap has an angular dependence with the azimuthal variable $\theta$ as 
it is sketched in Fig. \ref{fig1} and explicitly plotted in the upper-right picture of 
Fig. \ref{fig2}. With this configuration, the BEC 
is initially confined by a square potential in a region of angular dimension 
$-\alpha\leq\theta\leq 0$ along the ring. The scattering length is negative, allowing the 
cloud to form a soliton. The trapping well is connected to an optical 
lattice  of variable amplitude $A(\theta)$. 

To put the atomic soliton into motion, the value of the
scattering length is switched to zero in the confining region (indicated in orange color in the upper
plots of Fig. \ref{fig2}) and the condensate is emitted to the periodic structure 
(see Fig. \ref{fig2} a, b). The process is the same as in atom lasers based on control 
of the scattering length\cite{atomlaser}. Once the  condensate has left the trap 
and circulates through the ring, the initial confining potential can be removed and
$a$ is hold at a negative value along the whole cycle.

{\em Soliton acceleration.-}
In Fig. \ref{fig2}, we show a one dimensional version of our model. In this case,
we consider that $V_r\gg V_0$ and thus the cloud is tightly confined in the plane 
perpendicular to the ring. We have analyzed the case
of a set of $N\approx 10^5$ $^7$Li atoms with a negative scattering length. We must stress 
that our results are valid for other atomic species like $^{85}$Rb or $^{133}$Cs provided  
that $a$ is tuned to negative values. In our simulations we have used $a=-2nm$, which is 
easily accessible by adequate tuning of Feshbach resonances\cite{Feshbach}. We have 
chosen a ring such that the transverse width of the cloud is $w_\perp=0.6\mu m$. The longitudinal 
dimension of the cloud is $w\approx 10w_\perp$, thus for this cigar shape configurations,
transverse effects can be neglected in a first approximation.

As it can be seen in the upper plots of Fig. \ref{fig2}, the potential along the ring 
consists of two parts: a square trap of width $L$ (shaded region) and a periodic lattice 
connected to it. 
To put the cloud into motion from the trap of depth $V_0$ it is 
necessary that the energy of the self-interactions overcomes the potential energy of the well. 
This yields to the condition: $Na_{cr} \approx -4L_\perp V_0/V_r$\cite{atomlaser}. 
The part of the trapped cloud which overlaps the zone with $a<0$ feels an effective force 
due to the nonlinear interactions. If the value of $a$ is negative enough, this 
force overcomes the square potential and a robust atomic soliton\cite{solitons}
is emitted to the lattice. Thus, $a$ has a double effect in the condensate: first it acts
as a trigger for the motion of the soliton and in second place it avoids 
the spreading of the cloud during propagation\cite{murch} allowing major 
times for handling the cloud.

Once the condensate has been emitted from the trap, it is necessary  that the amplitude
$A(\theta)$ of the lattice increases along the ring in order to get acceleration of the center of mass of the 
cloud\cite{kartashov}. This can be appreciated in the lower pictures of Fig. \ref{fig2}. 
In Fig. \ref{fig2} a), we show the effect of a lattice with constant
amplitude. As it can be seen in the caption, no acceleration is obtained. In Fig. \ref{fig2} b) it 
can be appreciated the acceleration effect obtained with an exponentially modulated AM lattice, which is 
a configuration very suitable for experiments. We must stress that similar results can be obtained 
with any rapidly growing function, as we have tested numerically. This acceleration can be explained qualitatively 
with a variational method\cite{variational}. This well-know technique uses a Gaussian trial function 
to average the lagrangian density of Eq. \ref{GPE}. As a result, a second order 
ordinary differential equation is found for the center of mass of the condensate, located at an angle 
$\theta_0(t)$:

\begin{equation}
\label{Newton}
\ddot{\theta}_0 = \frac{-1}{4mr_0w}\frac{dI}{d\theta_0};
\end{equation}

where dots designate time derivatives, $m$ is the atomic mass, $w$ is the longitudinal with of the cloud,
$r_0$ the radius of the ring reservoir and $I$ is the following integral:

\begin{equation}
\label{I}
I = \frac{1}{\sqrt{\pi}}\int_{-\infty}^{\infty}\text{exp}
\left[-\frac{r_0^2(\theta-\theta_0)^2}{w^2}\right]V_0f(\theta)d\theta.
\end{equation}

Once the condensate has been emitted, if we assume that the frequency of the lattice is much 
faster than the oscillations of the condensate, we can approximate $f(\theta)=A(\theta)sin^2(\nu\theta)
\approx A(\theta)/2$ where the factor $1/2$ comes from averaging over the sine function that defines the 
lattice. In fact, our numerical simulations show that there is a critical value of the lattice period
below which the soliton is only affected by the modulation function $A(\theta)$ and not by the lattice. 
If we assume an exponential growth of the form $A(\theta)=\exp(\eta\theta)$, we can obtain 
a compact formula for the acceleration of the atomic soliton:

\begin{equation}
\label{acceleration}
\ddot{\theta}_0 = B\text{exp}\left(\eta\theta_0\right),
\end{equation}
being $B=\eta V_0e^{(\eta w/2r_0)^2}/(8mr_0^2)$. Eq. \eqref{acceleration} 
can be easily integrated for initial conditions $\theta_0(0)=\dot\theta_0(0)=0$
yielding to:

\begin{equation}
\label{position}
\theta_0(t) = \text{ln}\left[1-\text{tan}^2\left(\sqrt{\frac{B\eta}{2}}t\right)\right].
\end{equation}
Eqs. (\ref{acceleration}) and (\ref{position}) 
fit with the numerical simulation within an error of about $15\%$. Taking 
this factor into account, the agreement between the numerical simulation 
and the variational calculation is both qualitative and quantitative. 

We must finally stress that the lattice should be switched off and on before the soliton reaches the boundary in
order to avoid reflections. Repeating the process periodically, it
is possible to obtain acceleration cycles that increase the angular velocity of the beam.
For the experimental parameters given above we obtained a maximum variation 
of $30mm/s$ per cycle of about $1s$, without spreading of the cloud and using a static 
lattice. This suppose a a significant breakthrough comparing with previous proposals\cite{previous}.

\begin{figure}[htb]
{\centering \resizebox*{1\columnwidth}{!}{\includegraphics{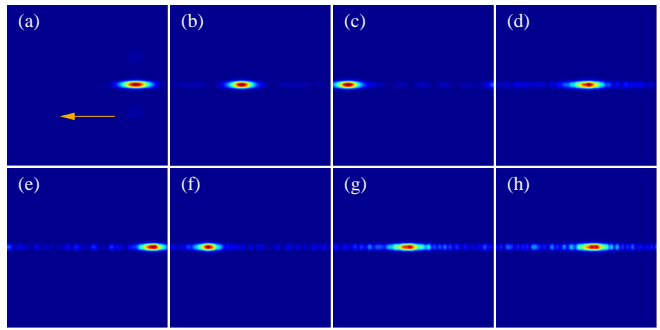}} \par}
\caption{[Color Online] Two-dimensional simulation showing the acceleration of 
a Bose-Einstein soliton in a rectilinear 2-D cavity in the presence of an AM optical lattice.
Each frame has been taken at regular intervals of time from $t=0$ to $t=2.5s$. 
The numerical values are the same as in Fig. \ref{fig2}.}
\label{fig3}
\end{figure}

In order to check the validity of the one-dimensional model, we display in Fig. \ref{fig3} 
a series of images from a two-dimensional simulation in a rectilinear trap.
The lattice in Fig. \ref{fig3} has twenty five lines in the space of the initial soliton 
size. For our simulations, this means a fringe size of 
$\approx 0.5\mu m$, which can be easily obtained experimentally.
The frames have been taken at the same intervals of time from $t=0$ to $t=2.5s$. 
The other parameters are the same as in Fig. \ref{fig2} b). 
As it can be appreciated from the pictures, the cloud keeps its shape in time, thus constituting a 
two-dimensional Bose-Einstein Soliton. This is achieved as a combination of the guiding ring, the lattice
and the nonlinear interactions. The periodic boundary conditions of the simulation allow to repeat the process for several cycles.

\begin{figure}[htb]
{\centering \resizebox*{1\columnwidth}{!}{\includegraphics{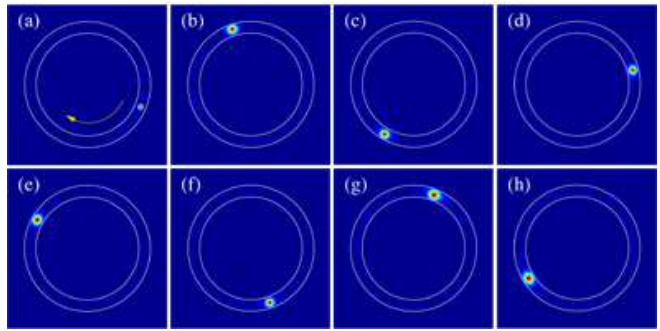}} \par}
\caption{[Color Online] Numerical simulation of the ring configuration. Each 
frame has been taken at regular time intervals from $t=0$ to $t=8000\nu_\perp^{-1}$. 
As it can be seen in the pictures, the cloud does not spread during the cycles.}
\label{fig4}
\end{figure}

In Fig. \ref{fig4} we show the full ring configuration. The numerical values are 
the same as in the previous figures. Each frame has been taken at regular time 
intervals from $t=0$ to $t=2550\nu_\perp^{-1}$. As it can be seen in the pictures, 
the cloud does not spread during the cycles. The combination of the guiding ring 
and the nonlinearity forms a {\em rotary soliton}\cite{rotary} which can be accelerated in 
each cycle. This opens new experimental possibilities with ultracold 
atom beams like particle colliders or precision atom interferometers.

\begin{figure}[htb]
{\centering \resizebox*{1\columnwidth}{!}{\includegraphics{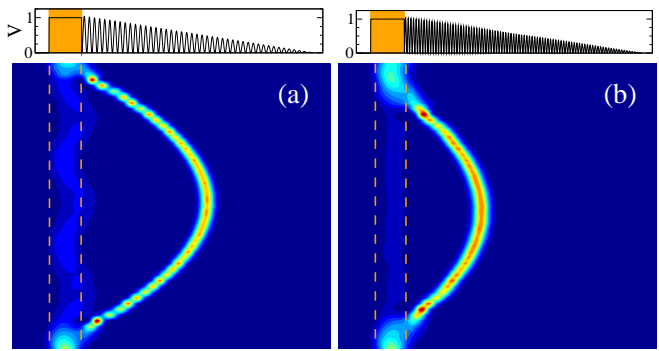}} \par}
\caption{[Color Online] Effect of an AM lattice with linear decrease of the
amplitude. Vertical axis is time from $t=0$ to $t=4500/\nu_\perp$ 
for $(a)$, to $t=4000/\nu_\perp$ for $(b)$. The rest of the parameters are the
same as in Fig. \ref{fig2}}
\label{fig5}
\end{figure}

{\em Soliton motion control.-}
The control of the motion of the Bose-Einstein Soliton, can be achieved by using 
AM or FM lattices. Fig. \ref{fig5} shows the effect of an AM lattice with linear 
decrease of the amplitude. Vertical axis is time from $t=0$ to $t=4500/\nu_\perp$ 
for $(a)$, and to $t=4000/\nu_\perp$ for $(b)$. The rest of the parameters are the
same as in Fig. \ref{fig2}. As it can be appreciated in the picture, the soliton  
turns back due to the decrease in the amplitude of the lattice. This effect could have potential use 
in the design of precise atomic interferometers. As it can be appreciated by 
comparing both images a) and b) in Fig. \ref{fig5}, the turning point can be precisely adjusted 
by simply changing the frequency of the lattice. 
                                 
\begin{figure}[htb]
{\centering \resizebox*{1\columnwidth}{!}{\includegraphics{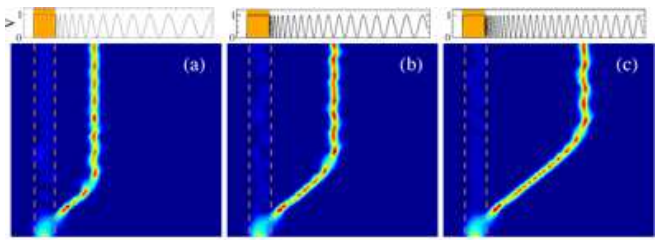}} \par}
\caption{[Color Online] Effect of a FM lattice. Its shape is shown for each case at 
the top of the pictures. Vertical axis is time from $t=0$ to $t=4000/\nu_\perp$; 
where $\nu_\perp$ is the radial trapping frequency. The rest of the parameters are the
same as in Fig. \ref{fig2}}
\label{fig6}
\end{figure}

On the other hand, if the frequency of the lattice decreases along the ring
of propagation, the soliton can be stopped at a given position. As it is shown in 
Fig. \ref{fig6} this allows a robust control of the atomic beam along the ring. We have 
verified that the solution can be generalized for a 2-D numerical simulation and different
modulations of the lattice. 
              
{\em Conclusions.-} In this work we have described a new scheme for obtaining acceleration
of ultracold atomic beams. Our system is capable of both increasing the angular velocity of
a BEC stored in a ring reservoir while avoiding its spreading. We have also discussed other
effects that can be obtained by using AM and FM lattices which allow a robust control of the
motion of the condensate. Our results ate fully testate in the frame of current experiments
with ultracold atomic gases.
                              
{\em Acknowledgments.-} This work was supported by Ministerio de Educaci\'on y Ciencia, Spain
(projects FIS2004-02466, BFM2003-02832, network
FIS2004-20188-E) and Xunta de Galicia (project PGIDIT04TIC383001PR).

\end{document}